\documentstyle[12pt,epsf]{article}
\newcommand{\beq}{\begin{equation}}
\newcommand{\eeq}{\end{equation}}
\newcommand{\beqa}{\begin{eqnarray}}
\newcommand{\eeqa}{\end{eqnarray}}

\begin{document}
 
\renewcommand{\thefootnote}{\arabic{footnote}}
 
\setcounter{footnote}{0}
 
\begin{flushright}
\end{flushright}
\vskip 5pt
\begin{center}
{\Large {\bf 
Mass and Scalar Cross-sections for Neutralino 
Dark Matter in Anomaly Mediated
Supersymmetry Breaking Model 
}}
\vskip 20pt
\renewcommand{\thefootnote}{\fnsymbol{footnote}}
 
{\sf Debasish Majumdar $^{a,\!\!}$
\footnote{E-mail address: debasish@theory.saha.ernet.in}}
\vskip 10pt
$^a${\it Theory Division, Saha Institute of Nuclear Physics,\\  
1/AF Bidhannagar, Kolkata 700064, India. }\\
\end{center} 
\vskip 5mm
{\small 
We have considered neutralino to be the lightest supersymmetric particle 
(LSP) in the framework of minimal Anomaly Mediated
Supersymmetric (mAMSB) model. We have studied variation of neutralino mass
with the supersymmetric parameters. Considering these neutralinos to be 
the candidates for weakly interacting massive particle (WIMP) or cold 
dark matter (CDM), we have calculated the neutralino 
nucleon scalar cross-sections and compared them with DAMA-NaI neutralino 
direct detection search results. From this study we observe that the 
mAMSB model results cannot explain the allowed region in WIMP mass and 
WIMP-nucleon scalar cross-section space obtained from annual modulation 
signature in DAMA-NaI experiment.
}
\vskip 1cm

Observational evidence like velocity curves of spiral galaxies, 
measurement of X-ray emissions of the cluster of galaxies, 
gravitational lensing and other
theoretical calculations strongly suggest the existence of enormous amount of
matter in the universe which is invisible to us. 
This invisible, nonluminous matter 
or `Dark Matter' constitute major fraction of the mass of the universe. 
While there may well be more than one type of dark matter, like baryonic
or non-baryonic, hot (relativistic) or cold (non-relativisitic), large 
scale structure calculations suggest non-relativistic massive particles (cold)
that were decoupled from the plasma when the universe was entering into 
the epoch of matter dominance, constitute a dominant fraction of dark matter. 
These cold dark matters should be made up of weakly interacting massive 
( $>$GeV) particles or WIMPs. 

The favourite candidate for the WIMPs for cold dark matter, are lightest 
supersymmetric particles or LSP.  As generally WIMPs are to be electrically
neutral, the widespread choice for LSP is neutralino. Neutralino is a neutral
Majorana particle and a superposition of fermionic super-partners of gauge and
Higgs bosons namely bino, wino and two neutral higgsinos. 

The direct detection of dark matter WIMPs can be realised by 
investigating their elastic scattering on the target nuclei of the detector. 
Hence for theoretical prediction of the WIMP detection rate it is 
important to calculate this cross-section for a choice of dark matter 
candidate. 

The WIMPs in the galactic halo are assumed to have an approximately 
Maxwellian velocity distribution in the galactic rest frame. But 
for an observer on the earth (for one who is moving with the velocity 
of the earth) one has to include earth's motion with respect to the halo. 
This has two components. One is sun's motion in galaxy and the
other is earth's motion with respect to the sun. The directionality 
of this latter motion changes over the year as earth rotates around the 
sun. This in turn induces an annual variation of the WIMP speed relative 
to earth (maximum when earth's rotational velocity adds up to the velocity 
of the sun and minimum when these velocities are in opposite directions).
This imparts an annual modulation of the direct detection rates of the WIMPs
constatnly encountered by earth \cite{drukier,freese}. 
DAMA- NaI detector at Gran Sasso in 
Italy \cite{bernabei} looks for this annual modulation signature 
in WIMPs direct search.   
They also have published allowed 3$\sigma$ confidence level contours 
for their dark matter search in $M_{W}$ - $\xi \sigma_p$ plane where 
$M_W$ is the WIMP mass, $\sigma_p$ is the WIMP -nucleon scattering cross 
section and $\xi = \rho_\chi / \rho_\ell$, $\rho_\chi$ being the local
dark matter density and $\rho_\ell$ is the halo density (taken to be 
0.3 GeV/cm$^3$).  

In this letter minimal Anomaly Mediated Supersymmetry Breaking 
(mAMSB) model  is cosidered for generating neutralino ($\chi$) LSP. It is 
our endeavour here to investigate the mass and scalar cross-sections 
for these neutralinos within the range of parameters given by mAMSB
model. We also investigate whether these mass and scalar cross-sections
can explain DAMA-NaI dark matter search results.

In this letter we consider the supersymmetric particle neutralino 
as a WIMP particle which is a possible
candidate for dark matter. Neutralino is the lowest mass eigenstate of
linear superposition of photino ($\tilde{\gamma}$), zino ($\tilde{Z}$),
and the two Higgsino states ($\tilde{H^0_1},\tilde{H^0_2}$) \cite{Haber}
and is written as 
\beq
\chi = a_1\tilde{\gamma}+ a_2\tilde{Z}+ a_3\tilde{H^0_1}+a_4\tilde{H^0_2} \, . 
\eeq 

The supersymmetric model we have chosen here is minimal Anomaly Mediated 
Supersymmetry Breaking (mAMSB) model \cite{randall,giudice}. In this model, the
observable sector (OS) and the hidden sector (HS) are in two distinct
3-brane separated by a finite bulk distance which is of the order of the 
compactification radius of a fifth compactified dimension. Unlike in ordinary
gravity-mediated supersymmetry breaking where the breaking is transmitted 
from HS to OS by tree level exchanges, in anomaly mediated supersymmetry
breaking model or AMSB model, the supersymmetry breaking
is transmitted through loop generated superconformal anomaly. An 
sparticle spectrum in this model is fixed by three parameters namely 
$m_{3/2}$ which is equal to the gravitino mass, the ratio $\tan\beta$ 
of the vacuum expectation values of two Higgs fields $H^0_1$ and $H^0_2$
and sign($\mu$) ($\mu$ being the Higgsino mass). The problem of tachyonic
sfermions in this model is remedied by cosidering a universal mass 
squared term $m_0^2$ that will make all sfermion masses positive. This is
minimal anomaly mediated supersymmetry breaking or mAMSB model. Therefore,
with $m_0$, we have four parameters in mAMSB model.

In Minimal Supersymmetric Standard Model (MSSM), the $\chi$ - nucleon
scattering cross-sections involve two independent parameters namely 
neutralino mass and squark mass. In AMSB model both the squark mass 
and the neutralino mass are determined by $m_{3/2}$ parameter. For 
phenomenological reasons, a universal mass term $m_0$ has to be added 
to the squark mass. However, the stability of SUSY potential \cite {ad}
constrains the $m_0 - m_{3/2}$ parameter space severly. Similar 
constraints cannot be obtained in MSSM. Hence our calculation has more
predictive power.  

Recently several authors had used this model for various calculations 
\cite{ad,sourov}. In the present calculations, with mAMSB model, we need to 
fix the range of parameters namely $m_{3/2}$, $\tan\beta$, $m_0$ and the 
sign of $\mu$. We have chosen Ref.\cite{ad} for this purpose. Datta et al in
Ref. \cite{ad} give the allowed parameter space for the mAMSB model 
for $\tan\beta =5$ in Fig 1. In Tables 1 and 2 of the same reference, 
Datta et al furnished some indicative values of the parameters (including
lower bounds on $m_0$ and upper bounds on $m_{3/2}$) from their calculations. 
Using them we have chosen the following ranges for $\tan\beta$ and $m_0$. 
\begin{eqnarray}
5 \leq & \tan\beta & \leq 15  \nonumber \\  
405~ \rm{GeV} \leq & m_{0} &\leq 905~ \rm{GeV}
\end{eqnarray}
The range of the parameter $m_{3/2}$ (in TeV) for each choice of $m_0$ 
is fixed by the allowed region in Fig 1 of Ref. \cite{ad}. The sign of
$\mu$ is taken to be negative.
 
We attempt here to calculate the scalar cross-section for different WIMP
mass. The WIMP here is neutralino dark matter. The neutralino mass 
and neutralino nucleon scalar scattering cross-sections are calculated in the 
framework of mAMSB model.

Bottino et al, who have done many pioneering work in dark matter calculations
gave an expression for neutralino-nucleon scalar scattering cross-section
\cite{bottino1,bottino2}. This expression (Eq. 2 of \cite{bottino1} and Eq. 1 of
\cite{bottino2}) reads as
\beq
\sigma_{\rm scalar} = 
\displaystyle\frac {8G_F^2} {\pi} M_Z^2 m_{red}^2 \left 
[ \frac {F_hI_h} {m_h^2} +  \frac {F_HI_H} {m_H^2} + 
\rm {squark}~~ \rm {exchange}~~ \rm {term}
\right ]^2
\eeq
The two terms in bracket in the above equations are $h$ and $H$ exchange terms
where $h$ and $H$ are CP even neutral Higgs bosons. 
\begin{eqnarray}
F_h &=& (-a_1 \sin\theta_W + a_2 \cos\theta_W)(a_3\sin\alpha + a_4 \cos\alpha)  
\nonumber \\
F_H &=& (-a_1 \sin\theta_W + a_2 \cos\theta_W)(a_3\cos\alpha + a_4 \sin\alpha)  
\nonumber \\
I_{h,H} & = & \displaystyle\sum_q k^{h,H}_q m_q \langle N | \bar{q} q | N 
\rangle
\end{eqnarray}
In the above $a_1$, $a_2$, $a_3$, $a_4$ are the coefficients as shown in 
Eq.(1), the angle $\alpha$ is the Higgs mixing angle that rotates 
$H_1^0$ and $H_2^0$ into $h$ and $H$, $m_h$ and $m_H$ are the masses of 
$h$ and $H$ respectively, $m_{\rm red}^2$ is the neutralino-nucleon
reduced mass. The last term in Eq. (3) can be written as \cite{bottino1}
\beq
I_{h,H} = k^{h,H}_{u-\rm{type}} g_u + k^{h,H}_{d-\rm{type}} g_d
\eeq
where $k^h_{u-\rm{type}} = \cos\alpha/\sin\beta$, 
$k^h_{d-\rm{type}} = -\sin\alpha/\cos\beta$, 
$k^H_{u-\rm{type}} = -\sin\alpha/\sin\beta$, 
$k^H_{d-\rm{type}} = -\cos\alpha/\cos\beta$ \cite{bottino1}.
The values of $g_u$ and $g_d$, in terms of pion-nucleon sigma term, 
$\sigma_{\pi N}$, are given in Ref. \cite{bottino1,bottino2}. In the present 
calculation, for simplicity, we deal only with Higgs mediated terms. 
Also it is pointed out in Ref. \cite{bottino1} that in general these terms
are largely dominant over the squark exchange term.

For the calculation of scalar cross-sections (Eq. 3-5) one has to compute
$a_1$, $a_2$, $a_3$, $a_4$, the Higgs mixing angle 
$\alpha$, the Higgs masses $m_h$ and $m_H$, the neutralino mass $m_\chi$.
These are calculated in the framework of mAMSB model for the range of 
supersymmetric parameters discussed above, using the code ISAJET 7.48
\cite{isajet}. We have checked that the neutralino is the LSP. 
In Fig. 1 we show the variation of neutralino mass 
$m_\chi$ with $\tan\beta$ for three sets of values of $m_0$ and $m_{3/2}$ 
with $\mu < 0$. Fig. 1 shows that there are no significant variations of 
$m_\chi$ with $\tan\beta$. In Fig. 2 we plot $m_0$ (in GeV) vs $m_\chi$
for three fixed sets of values of $m_{3/2}$ and $\tan\beta$. These three
sets are chosen by fixing $m_{3/2}$ at a lower value 31.5 TeV with 
three values of $\tan\beta$ namely 5.0, 10.0 and 15.0. It is observed from
Fig. 2 that although the plot corresponding to $\tan\beta = 5$ is rather 
flat the plots for $\tan\beta =10$ and 15.0 show some variations of neutralino
mass with $m_0$. More so the mean difference of $m_\chi$'s corresponding 
to $\tan\beta =5$ and $\tan\beta = 10$ is almost double than that 
of $\tan\beta =10$ and $\tan\beta = 15$. The variation of $m_\chi$ is 
therefore a collective effect of variation of the three mAMSB model parameters.

For calculation of neutralino-nucleon scalar cross-sections using 
Eq. (3-5), we express all the quantities in the units of GeV and the 
cross-sections are expressed in pico barn (pb). The values of  
$g_u$ and $g_d$ (in Eq. 5) are obtained from
\cite{bottino1} and they are $g_u = 123$ GeV, $g_d = 288$ GeV. 
The results of calculated scalar cross-sections, 
$\sigma_{\rm scalar}$ (in pb) for various netralino masses 
($m_\chi$ in GeV) are shown as scatter plot in Fig. 3. The particular 
pattern of this scatter plot for cross-section calculations using mAMSB model
is reflective of the shape of the allowed region for $m_0 - m_{3/2}$ parameter 
space for mAMSB model as shown in Fig 1. of Datta et al \cite {ad}.  

In order to study whether these results can represent the DAMA-NaI results 
stated above, we also show in Fig. 3, the results obtained from DAMA-NaI
experiment. They are shown as contour plots in Fig. 3.
The contours in Fig. 3 (continuous plots) are obtained from Fig. 4b 
of Ref. \cite{bernabei} where they have 
given 3$\sigma$ C.L. allowed regions for experiments in
four different yearly cycles (DAMA-NaI-1 to DAMA-NaI-4, with total 
statistics of 57986 kg.day) alongwith DAMA-NaI-0 results \cite{dama2}. The 
plots suggest that although the neutralino mass range falls within 
that obtained from DAMA-NaI experiments but the cross-sections are about 
an order of magnitude below than the DAMA contour.  An attempt to find out
$\xi = \rho_\chi/\rho_\ell$ where $\rho_\ell=0.3$ GeV/cm$^3$ \cite{bernabei} 
using the relation $\xi = (\xi\sigma)_{DAMA}/\sigma_{calc.}$, leads to values
of  $\rho_\chi \sim 3$ GeV/cm$^3$ which is much more than the halo density 
$\rho_\ell$ and thus is unphysical.

To conclude, we have considered minimal Anomaly Mediated Supersymmetry 
Breaking model for studying the variation of neutralino mass with the 
supersymmetric parameters, namely $\tan\beta$, $m_0$ and $m_{3/2}$. We have 
calculated the netralino nucleon scalar cross-sections and studied their 
variations with neutralino mass. Considering neutralino to be cold dark matter
candidate, we went on to compare the results with those obtained from 
DAMA-NaI detection of annual modulation of dark matter signal. 
It appears that the mAMSB model calculations cannot reproduce DAMA-NaI 
experimental findings for WIMPs. Although earlier results from calculations 
within MSSM frame-work can represent the DAMA results to some extent, for mAMSB 
model it is not so and it is due to the fact that the parameter space 
of mAMSB model is somewhat different from MSSM. 
As discussed earlier, mAMSB model
has only four parameters and $m_0 - m_{3/2}$ parameter space is severly 
constrained. This affects the calculation of the coefficients $a_1$, 
$a_2$ etc. and the value of $\alpha$. The scalar cross-sections thus 
obtained using Eqs. (3 - 5) for different $m_\chi$'s are therefore different 
from those obtained in MSSM.     
We repeat our calculations  within the framework of mAMSB model  
with different $g_u$, $g_d$ values given in Table 2 of a more recent 
work by Bottino et al \cite{bottino2}. We find these 
cross-sections fall by orders
of magnitudes from those calculated using \cite{bottino1}. 
These results are further off from the 
results of DAMA-NaI experiments. We have plotted these results 
in Fig. 4 for demonstration. The calculation of neutralino relic density in 
this model requires the calculations of neutralino-neutralino annihilation 
cross-sections and these will constitute our future work.

The author is grateful to Anindya Dutta for introducing him to the 
computer code ISAJET and for many valuable discussions. The author also 
thanks Amitava Datta for some useful discussions.

\newpage
\begin{center}
{\bf Figure Captions}
\end{center}

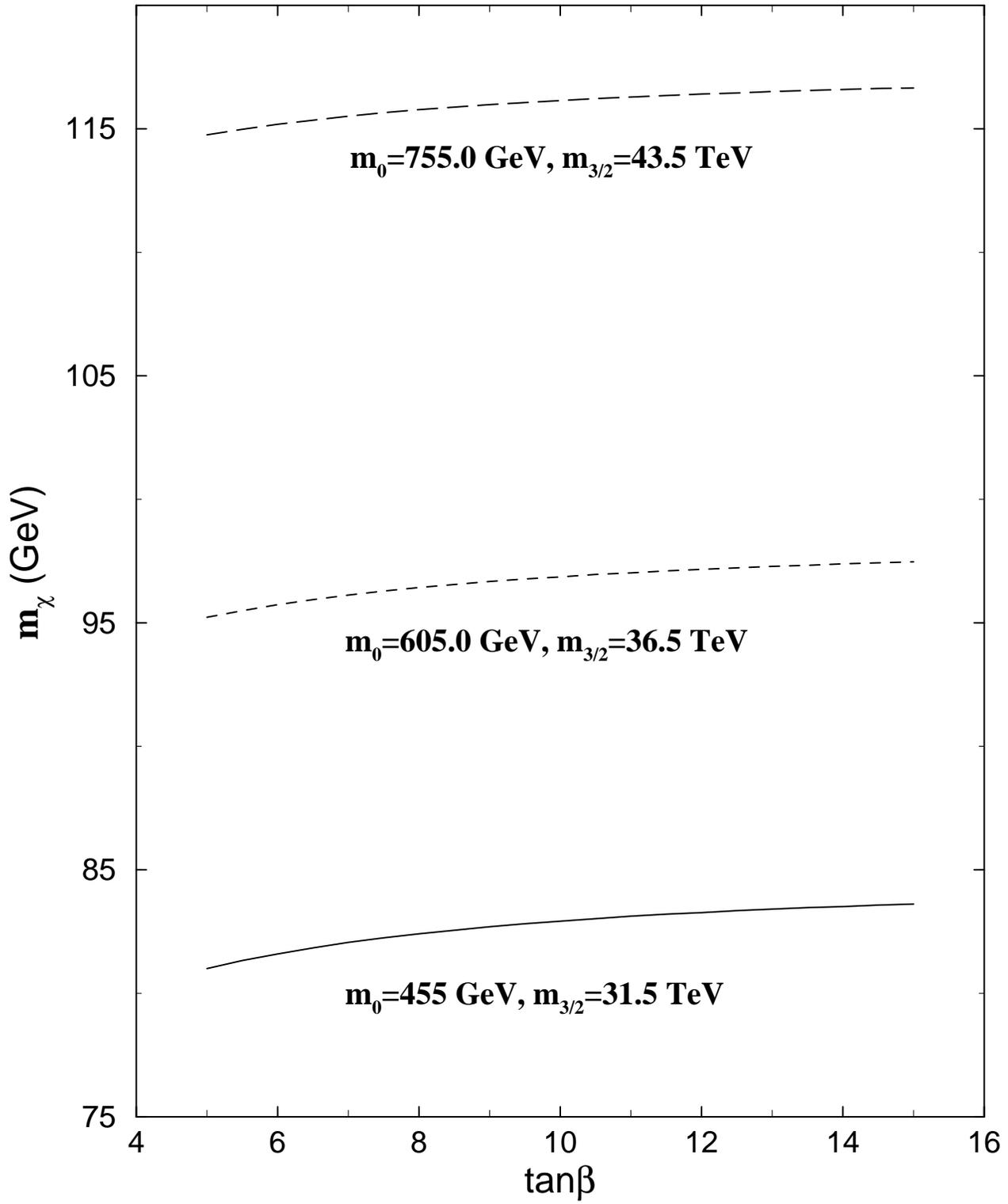
\noindent {\bf Fig. 1} Variation of neutralino mass $m_\chi$ (in GeV)
with $\tan\beta$ in
mAMSB model for three sets of values of $m_0$ (in GeV) and $m_{3/2}$ in TeV.
$\mu < 0$.

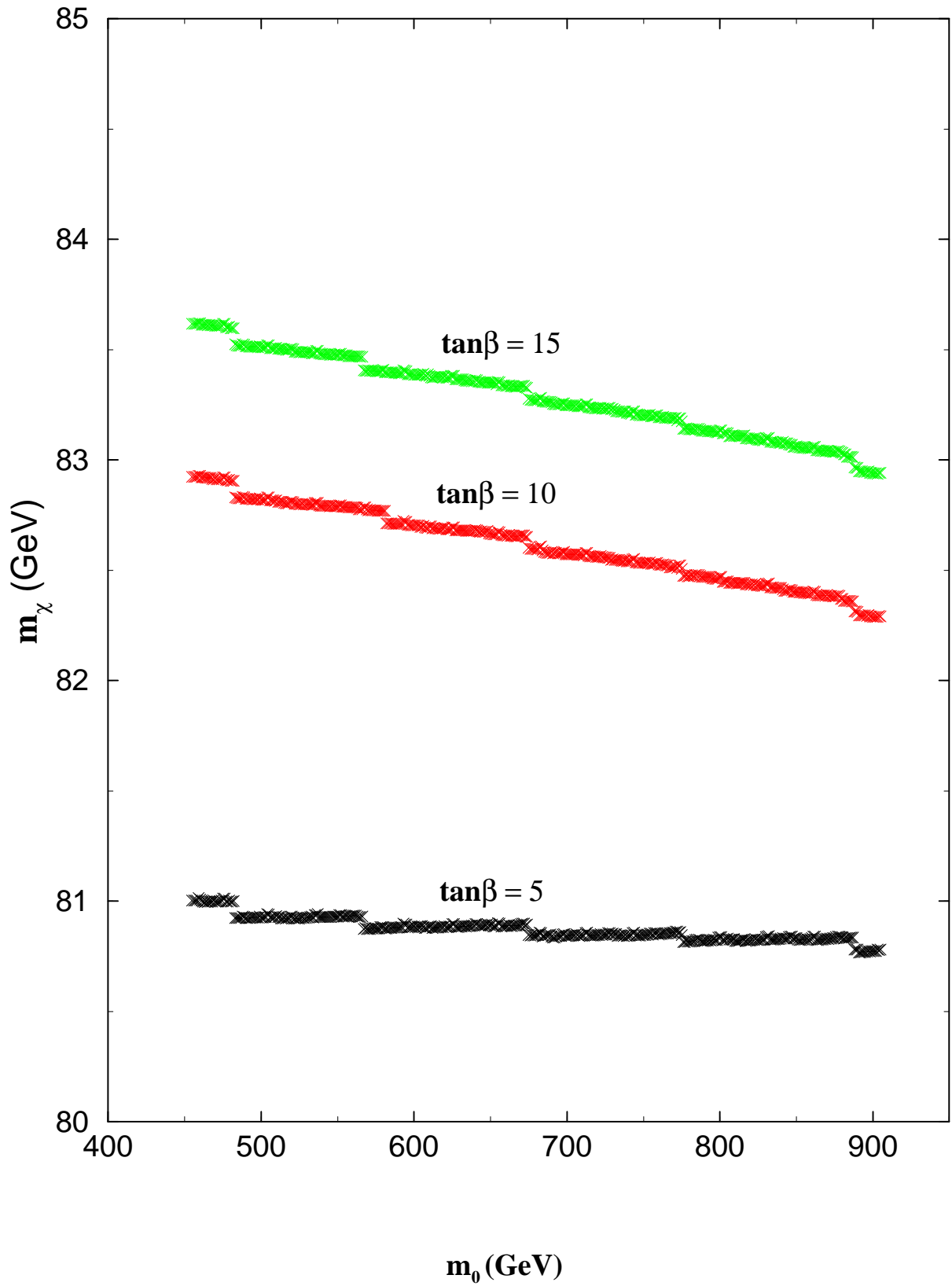
\noindent {\bf Fig. 2} Variation neutralino mass $m_\chi$ (in GeV) with 
$m_0$ in GeV for three sets of values of $m_{3/2}$ and $\tan\beta$. See 
text for detail.

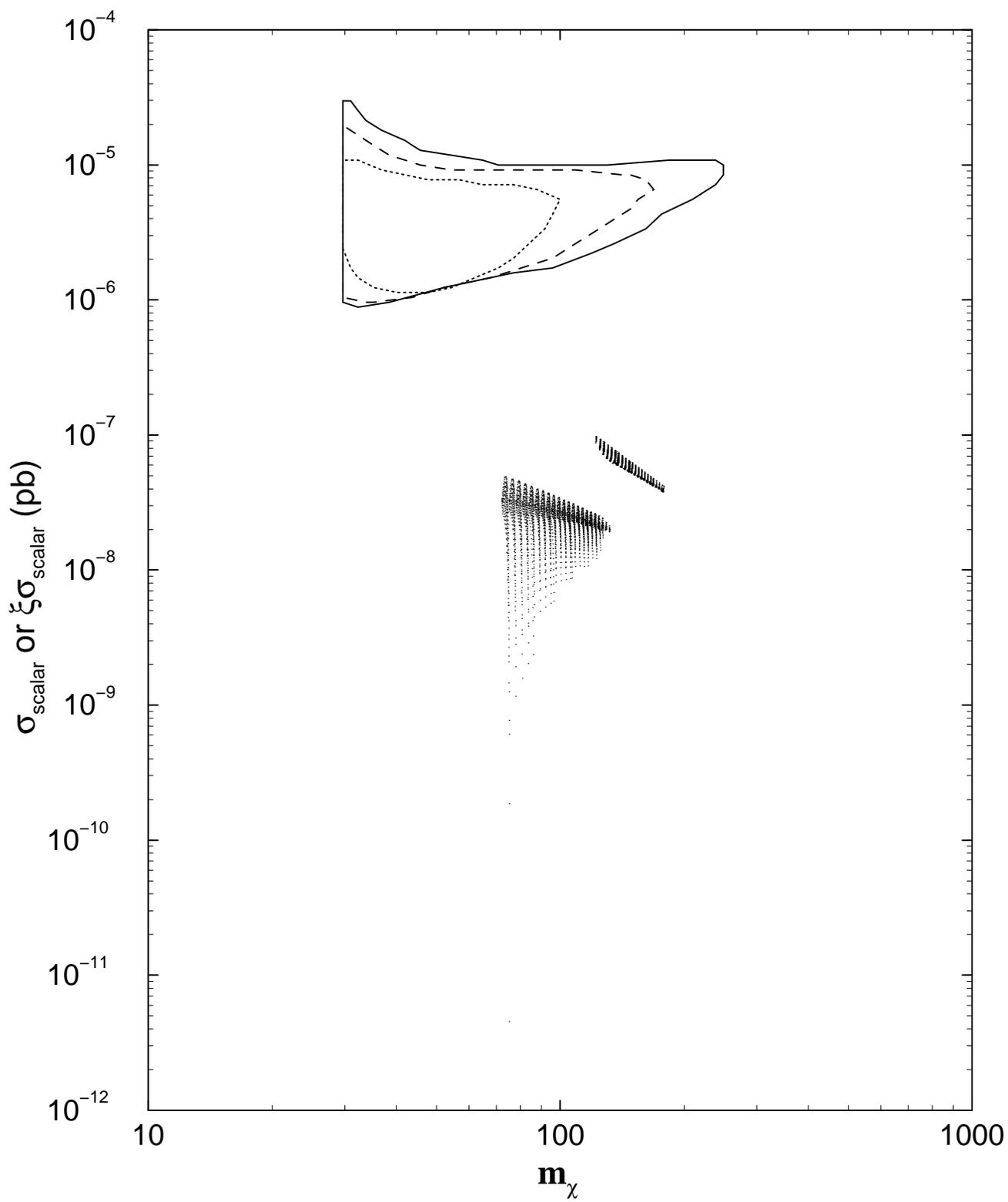
\noindent {\bf Fig. 3} Variation of neutralino nucleon scalar cross-section 
$\sigma_{\rm scalar}$ in pb (pico barn) with neutralino mass
$m_\chi$ in (GeV) (scatter plot)
and their comparison with DAMA-NaI results (see text 
for details). For DAMA-NaI results (closed contours) $\xi \sigma$ 
in pb is plotted along the
Y-axis while  for the results of present calculations 
$\sigma_{\rm scalar}$ in pb are plotted along Y-axis.

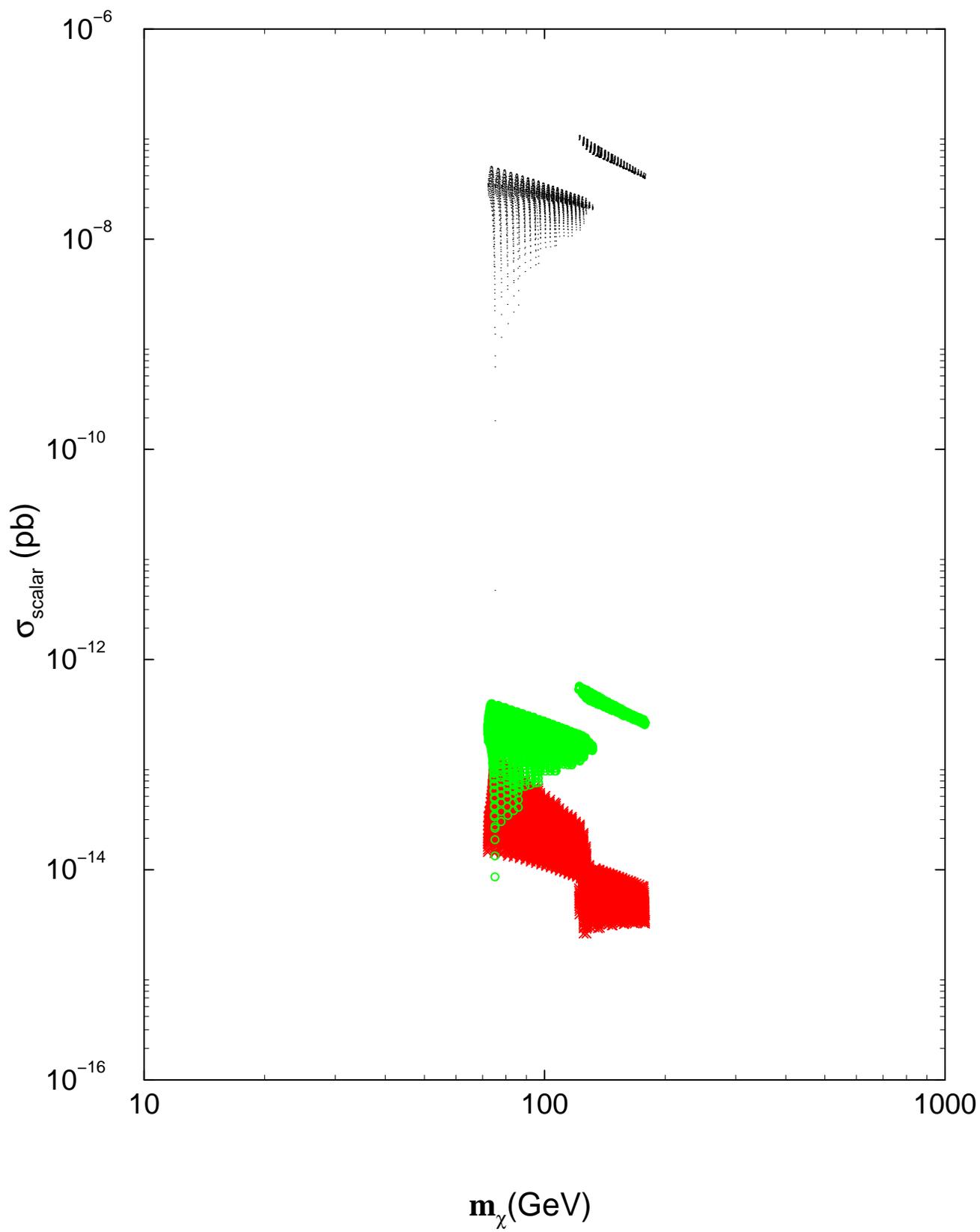
\noindent {\bf Fig. 4} Comparison of  cross-section 
($\sigma_{\rm scalar}$ in pb) calculations for different sets 
of values for $g_u$ and $g_d$. The symbols 'X' and circle corresponds to 
data sets (a) and (b) respectively in Table 2 of [10], the symbol '.' (dots) 
are for the same sets as in Fig. 3.
\end{document}